\documentclass[english]{article}
\usepackage[T1]{fontenc}
\usepackage[latin9]{inputenc}
\usepackage{geometry}
\usepackage{float}
\usepackage{amsmath}
\usepackage{graphicx}
\usepackage[authoryear]{natbib}

\makeatletter

\floatstyle{ruled}
\newfloat{algorithm}{tbp}{loa}
\providecommand{\algorithmname}{Algorithm}
\floatname{algorithm}{\protect\algorithmname}

\date{}

\@ifundefined{showcaptionsetup}{}{%
 \PassOptionsToPackage{caption=false}{subfig}}
\usepackage{subfig}
\makeatother

\usepackage{babel}
\usepackage{listings}

\begin{document}

\title{Offset Sampling Improves Deep Learning based Accelerated MRI Reconstructions
by Exploiting Symmetry}

\author{Aaron Defazio \\
Facebook AI Research NY, fastMRI team}
\maketitle
\begin{abstract}
Deep learning approaches to accelerated MRI take a matrix of sampled
Fourier-space lines as input and produce a spatial image as output.
In this work we show that by careful choice of the offset used in
the sampling procedure, the symmetries in k-space can be better exploited,
producing higher quality reconstructions than given by standard equally-spaced
samples or randomized samples motivated by compressed sensing.
\end{abstract}

\section{Problem setup}

Deep learning based accelerated MRI reconstruction methods are typically
trained using supervised machine learning. Our work uses the recently
released fastMRI multi-coil dataset \citep{zbontar2018fastMRI}, which
consists of a set of raw $k$-space data $x^{(j)}$ in multi-dimensional
arrays $:n_{c}\times h\times w$. From this data a ground-truth image
$m^{(j)}$ is produced, using a direct IFFT followed by a standard
sum-of-squares approach to combing coil images.

These training pairs are then used to train a black box model $B$
(typically a neural network), which maps from raw subsampled k-space
tensors to $h\times w$ spatial images. For instance, the training
loss for data-point $j$ is given by:
\[
L^{(j)}(\phi)=SSIM\left(B_{\phi}\left(M\left(x^{(j)}\right)\right),\,m^{(j)}\right),
\]
where $M$ is a masking function that zeros out a fraction of the
$k$-space lines, simulating a subsampling process. This work is primarily
concerned with the choice of masking function $M$. Due to the physics
of the MRI acquisition process, the mask is the same for each coil,
and it is most efficient to capture full rows or columns at a time.
For the purposes of discussion we will consider the simplified case
of a 1D FFT, with a single channel, as this captures the important
aspects of the sampling processes. The same sampling mask may be extended
to 2D images and multiple coil channels by duplicating the mask for
each coil, and extending the mask along the additional spatial dimension
to capture full lines of k-space.

\section{1D FFT Notation}

We call the \textit{image-space} image the magnetization image $x(n)$,
of width $N$, and it's Fourier transform the \textit{k-space} image
$X(k)$, also of width N. The (1D) DFT is defined as:
\[
X(k)=\sum_{n=0}^{N-1}x(n)\exp\left(-2\pi ikn/N\right)
\]
and the IFT is:
\[
x(n)=\sum_{k=0}^{N-1}X(k)\exp\left(2\pi ikn/N\right).
\]

\section{Equispaced Sampling}

A naive application of equispaced sampling is to form the masked k-space
$Y$ from the unmasked space $X$ via:
\[
Y(k)=\begin{cases}
X(k) & k\mod R=0\\
0 & \text{otherwise}
\end{cases}
\]
Where $R$ is the subsampling factor, or acceleration factor.

\begin{figure}
\begin{centering}
\subfloat[Image space, with grey as 0, real \& imaginary components shown]{\includegraphics[width=5cm]{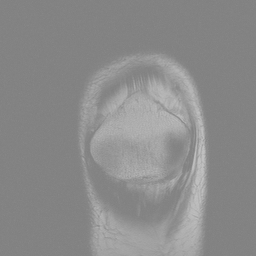}\hspace{1em}\includegraphics[width=5cm]{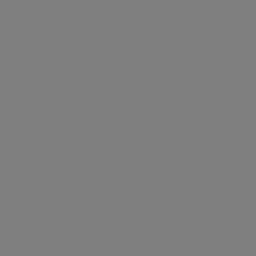}}
\par\end{centering}
\begin{centering}
\subfloat[\label{fig:Naive-R=00003D4-IFFT}Shift-0 IFFT real \& imaginary components]{\includegraphics[width=5cm]{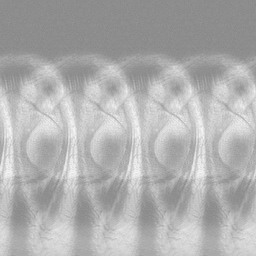}\hspace{1em}\includegraphics[width=5cm]{figures/illustration_offset0_imag}}
\par\end{centering}
\begin{centering}
\subfloat[Shift-1 IFFT real \& imaginary components]{\includegraphics[width=5cm]{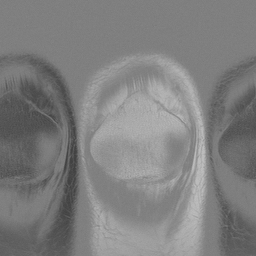}\hspace{1em}\includegraphics[width=5cm]{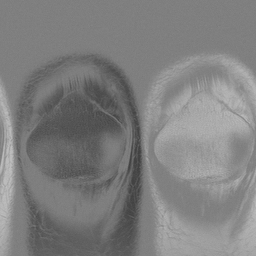}}
\par\end{centering}
\begin{centering}
\subfloat[Original (left) versus a shift-1 reconstruction (right) from clamping
negative values]{\includegraphics[width=5cm]{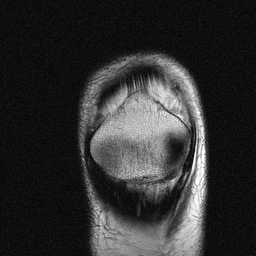}\hspace{1em}\includegraphics[width=5cm]{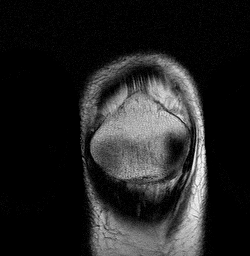}}
\par\end{centering}
\caption{\label{fig:ideal-equi}IFFTs in an idealized case, where the input
is real valued and takes up less than half of the horizontal field-of-view.}
\end{figure}

The behavior of the masking operation is interesting. It results in
the IFFT $y(m)=IFFT(Y(k))$ containing $R$ copies of $x(m)$, the
true magnetization image, each offset by $N/R$ from the last, for
width $N$. 
\[
y(m)=\frac{1}{R}\sum_{r=0}^{R-1}x(m+rN/R)
\]
by convention indexing of images is modulo the image size, so these
duplicates wrap at the image boundaries (Figure \ref{fig:Naive-R=00003D4-IFFT}).
We focus on the case where the k-space width is a multiple of R, as
otherwise the offsets $rN/R$ do not align with pixel boundaries.
This is known as aliasing in the Fourier literature, and in particular
this masking operation can be considered the composition of a DOWNSAMPLE
operation followed by a STRETCH operation. A model must thus ``decode''
this image by disentangling the R pixels that sum at each location.

We can provide a potentially more useful image to the model by considering
instead a sampling mask shifted by a single pixel:
\[
Y(k)=\begin{cases}
X(k) & k\mod R=1\\
0 & \text{otherwise}
\end{cases}.
\]
A remarkable property of this mask is that the shifted copies of the
image are now no longer in-phase but rather each differs by a phase
of $2\pi/R$ radians:
\[
y(m)=\frac{1}{R}\sum_{r=0}^{R-1}x(m+rN/R)\exp\left(2\pi r/R\right)
\]

This offset-1 sampling mask has a number of advantages, particularly
for the $R=4$ case which is of particular practical interest as it
provides a good balance of acceleration and reconstruction quality.
The 4 phases are real-positive, imaginary-positive, real-negative
and imaginary-negative.

If the magnetization image is also real-valued, then the real channel
of the IFFT will contain only 2 aliased images, separated by half
the image width, one positive and one equal but negated. This is effectively
half the overlap from what occurs in the offset-0 case. In the case
where the anatomy only takes up half the width, the two will be completely
separated, resulting in a perfect reconstruction modulo noise just
by clamping negative values to 0. The anatomy taking up half the width
or less is not uncommon as many coils are sensitive to signals in
only a portion of the field of view in common coil designs.

This shifted equispaced pattern is particularly advantageous if the
signal is real-valued, since there will be less overlap in the aliased
copies, as shown in Figure \ref{fig:ideal-equi}. Acceleration from
exploiting the real-valued nature of a signal is known as partial
Fourier \citep{partial-fourier,haacke1991fast,partial-fourier-eval93}
in the MRI literature. The offset-1 sampling scheme is able to automatically
make use of conjugate symmetry when possible.

\section{Phase shift derivation}

The above result is easiest to show by working backwards. We start
with image space:
\[
y(m)=\frac{1}{R}\sum_{r=0}^{R-1}x(m+rN/R)\exp\left(-2\pi ir/R\right),
\]
then we take the FFT:
\[
Y(k)=\frac{1}{R}\sum_{r=0}^{R-1}\exp\left(-2\pi ir/R\right)\sum_{n=0}^{N-1}x(n+rN/R)\exp\left(-2\pi ikn/N\right).
\]
The FFT of a shifted image is equal to multiplication by a linear-phase
term:
\[
\mathcal{F}(x(n+a))(k)=\exp(2\pi iak/N)X(k),
\]
so 
\begin{align*}
Y(k) & =\frac{1}{R}\sum_{r=0}^{R-1}\exp\left(-2\pi ir/R\right)\exp(2\pi i\left(rN/R\right)k/N)\left[\sum_{n=0}^{N-1}x(n)\exp\left(-2\pi ikn/N\right)\right]\\
 & =\frac{1}{R}\sum_{r=0}^{R-1}\exp\left(-2\pi ir/R\right)\exp(2\pi i\left(rN/R\right)k/N)X(k)\\
 & =X(k)\cdot\left[\frac{1}{R}\sum_{r=0}^{R-1}\exp\left(-2\pi ir/R+2\pi i\left(r/R\right)k\right)\right]\\
 & =X(k)\cdot\left[\frac{1}{R}\sum_{r=0}^{R-1}\exp\left(2\pi i\left(r/R\right)\left(k-1\right)\right)\right].
\end{align*}

The sum can be computed using sum-of-a-gemetric-series machinery.
Let $\omega=\exp\left(2\pi i/R\right)$ be the first root of unity.
Then using the formula:
\[
\sum_{r=0}^{R-1}\omega^{r(k-1)}=\frac{\omega^{R(k-1)}-1}{\omega^{k-1}-1}=\frac{1-1}{\omega^{k-1}-1}.
\]
So as long as the denominator is not zero, the sum is zero. The denominator
is zero when $k-1\mod R=0$ as the powers of unity wrap around at
R, and in that case we have 
\[
\sum_{r=0}^{R-1}\omega^{r(k-1)\mod R}=\sum_{r=0}^{R-1}\omega^{0}=R.
\]
The result is:
\[
Y(k)=\frac{1}{R}X(k)\cdot\begin{cases}
R & \left(k-1\right)\mod R=0\\
0 & \text{otherwise}
\end{cases}.
\]

\section*{An alternative view: exploiting conjugate symmetry}

The above mathematical treatment does not shed much light on \emph{why}
using an alternative offset results in more information being retained.
In this section we give a more direct argument. Consider the DFT as
stored in memory. We consider a simple 1D case of width 12. The Nyquist
frequency is $f=$1/6. The standard frequency layout in memory is
the following:
\[
\left[\begin{array}{cccccccccccc}
0f & 1f & 2f & 3f & 4f & 5f & -6f & -5f & -4f & -3f & -2f & -f\end{array}\right].
\]
When using the regular equispaced sampling, we keep the following
frequencies:
\[
\left[\begin{array}{cccccccccccc}
0f & 0 & 0 & 0 & 4f & 0 & 0 & 0 & -4f & 0 & 0 & 0\end{array}\right],
\]
Now recall that for real images, the k-space component for a frequency
is equal to that of the conjugate of its negative frequency. This
is known as the conjugate symmetry property of the real-value FFT.
So using an offset-0 sampling pattern actually keeps 2 copies of the
same frequency, giving us redundant information.

In contrast, the offset-1 pattern gives:
\[
\left[\begin{array}{cccccccccccc}
0 & 1f & 0 & 0 & 0 & 5f & 0 & 0 & 0 & -3f & 0 & 0\end{array}\right]
\]
So offset-1 sampling retains more frequencies, once conjugate symmetry
is taken into account. 

\subsection*{FFT shift}

It is common to store and manipulate FFT data in shifted configuration,
as given by the ``fftshift'' operation in Matlab or NumPy. This
shift operation moves the low frequencies to the center of k-space,
which is desirable for visualization purposes. The sampling procedure
described above must be performed on unshifted data as the sampling
mask will otherwise have an incorrect offset pattern. To avoid unnecessary
shifting of the full data matrix, the mask can be computed assuming
unshifted data, then the mask itself can be shifted to match the data's
layout.

\subsection*{Handling irregularly sized inputs.}

The above theory relies on the k-space image being an exact multiple
of the acceleration factor. When using non-multiples of the acceleration
factor the shifted copies occur at fractional pixel locations which
results in blurring. Although we strongly recommend performing scans
with the width in the phase-encoding direction a multiple of the acceleration
factor, it is not always practical. 

To get the correct pattern for image widths that are not a multiple
of the acceleration factor, the positive and negative parts of the
mask should be calculated separately. In the case of acceleration
factor 4, the offset for the negative part should be 2, counting backward
from the end (i.e. frequencies $-3f$, $-7f$, $\dots$). Numpy code
that produces correct masks is given in Algorithm \ref{alg:python-impl}.

\subsection*{Handling phase shifts}

Most practical MRI acquisitions exhibit a non-uniform and non-zero
phase, so the idealized case of all real-valued images is far from
the norm. Nevertheless, the described sampling mask still yields additional
information when the magnetization image is close to uniform in phase.
Since the masked data is input to a learned black-box model, this
information can potentially be used by the model to provide improved
reconstructions, without requiring any explicit partial Fourier techniques.

\begin{algorithm}
\begin{lstlisting}
offset_pos = 1             
offset_neg = 2
poslen = (n+1)//2
neglen = n - poslen
mask_positive = np.zeros(poslen, dtype=np.float32)
mask_negative = np.zeros(neglen, dtype=np.float32)

mask_positive[offset_pos::acceleration] = 1
mask_negative[offset_neg::acceleration] = 1
mask_negative = np.flip(mask_negative)

mask = np.concatenate((mask_positive, mask_negative))
\end{lstlisting}

\caption{\label{alg:python-impl}Python code that correctly produces an offset-1
sampling mask for irregularly sized inputs}

\end{algorithm}

\begin{figure}
\begin{centering}
\includegraphics[width=0.5\columnwidth]{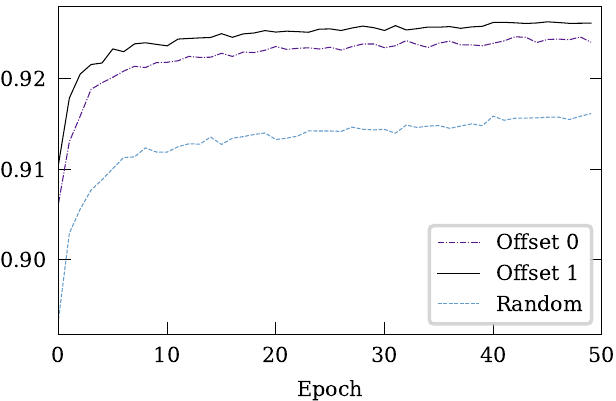}
\par\end{centering}
\caption{\label{fig:loss-comparison}Comparison of offset sampling for 4x accelerated
knee MRI reconstruction, showing an improvement in SSIM from 0.924
to 0.926.}
\end{figure}

\section{Experiments}

\label{sec:experiments}We compared the sampling approaches by training
a deep-learning-based system that takes a set of masked k-space channel
images as input and produces a spatial image as output. The training
procedure followed the setup and multi-coil dataset from \citet{zbontar2018fastMRI}.
We used a cascade of four U-Nets \citep{unet2015} following \citet{deepscascade-mri2018},
although we found our results were robust to the particular reconstruction
model used. The output of each U-Net is projected onto the known k-space
values (via a FFT-set-IFFT operation) before the application of the
next U-net. We used 4x subsampling for each mask, with the 16 lowest-frequency
lines also added to each mask via an OR operation.

Training consisted of 50 epochs of ADAM method with learning rate
$0.0003$, $\beta_{1}=0.9,$ $\beta_{2}=0.999$, batch-size 8, and
no weight decay, data augmentation or other regularization. The use
of offset 1 sampling significantly improves the validation SSIM, as
shown in Figure \ref{fig:loss-comparison}, and is superior to both
random sampling and offset 0 sampling. A set of example reconstructions
from the validation set is given in the Appendix, Figures \ref{fig:examples}
to \ref{fig:examples-1-1-1}. Both the standard offset 0 and the proposed
offset 1 methods produce good results. However, the offset 1 approach
produces sharper reconstructions, with more accurate low-level details.
We have highlighted regions where the differences are most prominent,
but smaller differences are visible through-out. The differences are
stronger when the signal-to-noise ratio is low.

\section{Discussion}

The use of randomized sampling for accelerated MRI has been a focus
of significant research \citep{sparseMRI2007} as an application of
the theory of compressed sensing. It may be surprising then that our
empirical results show a significant advantage to equispaced sampling.
We attribute this to a violation of the assumptions of compressed
sensing when applied to MRI images. For compressed sensing theory
to apply, there must exist some basis in which the image is near-sparse.
For the proton density scans with and without fat-suppression used
in the fastMRI dataset, there is significant textural detail which
are not sparse in image space or common wavelet bases. Random sampling
is much better suited to other MRI scan types such as vascular MRI
\citep{vascular-cs2018}, where the images are naturally sparse.

When equispaced sampling is used, the reconstruction problem becomes
a problem of disentangling multiple overlapping copies of the anatomy
when considered in image-space. This is a task that convolutional
neural networks are well-suited for, as the information about each
pixel location in the final image is localized to the neighborhood
of $R$ (for acceleration factor $R$) spots in the IFFT image. The
cascaded U-Net architecture uses an alternating sequence of local
operations (U-Nets) combined with non-local operations (projection
onto known k-spaces lines) to reconstruction the image.

In contrast, when a random sampling pattern has been used the information
needed to recover each location is spread throughout the image. Convolutional
neural networks use small kernels that are not well-suited to aggregating
information that is so spread out.

The use of offset sampling may not result in improved reconstruction
quality when classical accelerated imaging \citep{parallel-review2012}
approaches such as SENSE \citep{sense1999} or GRAPPA \citep{grappa2002}
are used, since they do not take advantage of the conjugate symmetry
present.

\section*{Conclusion}

The offset-1 approach shows a clear empirical advantage over other
fixed sampling approaches and is well supported by theory. The interpretation
of the approach as capturing non-redundant lines in k-space between
the positive and negative halves relates it to past work that exploits
conjugate symmetry such as partial Fourier approaches \citep{partial-fourier}.
While partial Fourier approaches typically fully capture the positive
half and partially capture the negative half, in the accelerated setting
it's possible to capture a subset of lines from both halves, as long
as the frequencies differ. The offset-1 sampling mask can be seen
as an extension of this idea to accelerated deep-learning-based reconstructions.

Our results suggest an interesting avenue for research: if the coil
design is modified to keep the spatial region captured by each coil
to below half the field-of-view, then it may be possible to further
improve reconstructions. Additionally, it may be possible to design
the coil sensitivity such that the phase varies in a way that ensures
the overlap between the multiple aliased copies occurs out-of-phase
at all spatial locations.

\section*{Acknowledgements}

This work was made possible through close collaboration with the fastMRI
team at Facebook AI research, including Nafissa Yakubova, Anuroop
Sriram, Jure Zbontar, Larry Zitnick, Mark Tygert, Suvrat Bhooshan
and Tullie Murrell, and our collaborators at NYU Langone Health on
the fastMRI project \citep{zbontar2018fastMRI}, with special thanks
to Florian Knoll, Matthew Muckley, Daniel Sodickson and Michael Recht.

\bibliographystyle{plainnat}
\bibliography{offset_masking}

\appendix
\begin{figure}
\subfloat[Fully sampled ground truth]{\includegraphics[width=0.49\columnwidth]{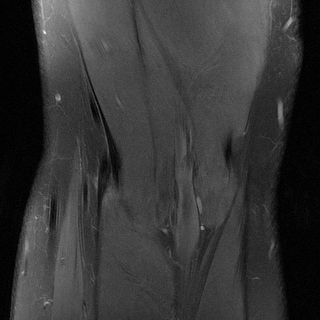}

}\subfloat[Offset 0 sampling ]{\includegraphics[width=0.49\columnwidth]{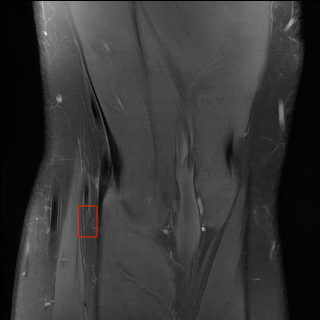}

}

\subfloat[Offset 1 sampling]{\includegraphics[width=0.49\columnwidth]{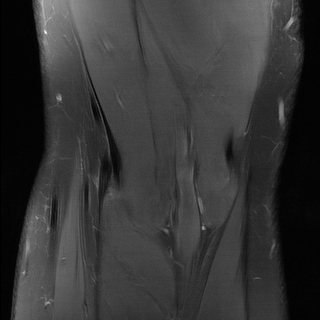}

}\subfloat[Random sampling ]{\includegraphics[width=0.49\columnwidth]{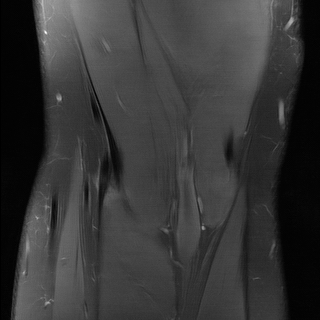}

}\caption{\label{fig:examples}Comparison of reconstructions from deep-learning
based systems trained using identical model architectures, and differing
sampling procedures, with areas of significant difference highlighted.}
\end{figure}

\begin{figure}
\subfloat[Fully sampled ground truth]{\includegraphics[width=0.49\columnwidth]{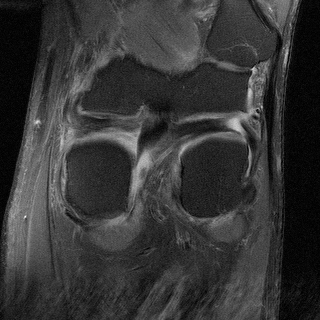}

}\subfloat[Offset 0 sampling ]{'\includegraphics[width=0.49\columnwidth]{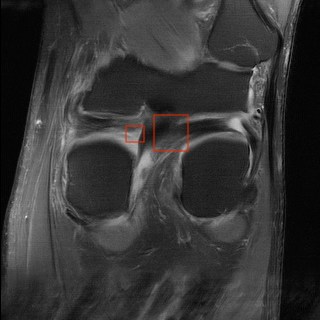}

}

\subfloat[Offset 1 sampling]{\includegraphics[width=0.49\columnwidth]{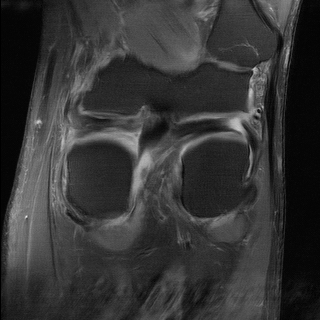}

}\subfloat[Random sampling ]{\includegraphics[width=0.49\columnwidth]{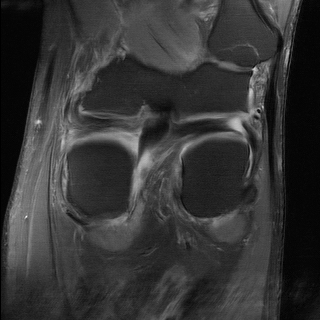}

}\caption{\label{fig:examples-1}In this example the offset-0 sampling mask
loses detail in the smaller highlighted region. The detail differs
significantly in the larger highlighted region, where the ground truth
is also unclear. Images are best viewed at full resolution on a high-brightness
monitor.}
\end{figure}

\begin{figure}
\subfloat[Fully sampled ground truth]{\includegraphics[width=0.49\columnwidth]{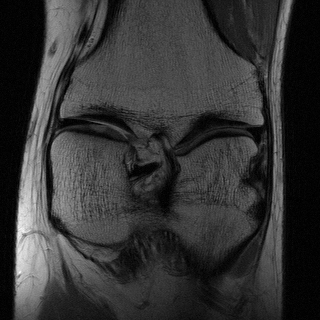}

}\subfloat[Offset 0 sampling ]{'\includegraphics[width=0.49\columnwidth]{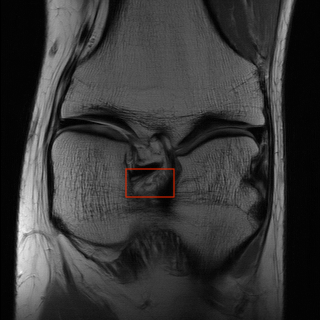}

}

\subfloat[Offset 1 sampling]{\includegraphics[width=0.49\columnwidth]{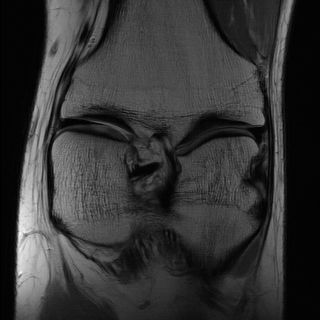}

}\subfloat[Random sampling ]{\includegraphics[width=0.49\columnwidth]{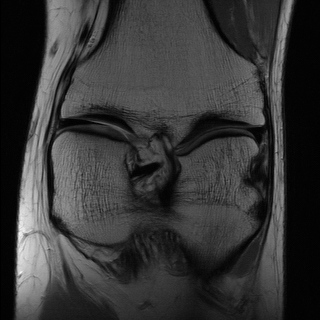}

}\caption{\label{fig:examples-1-1}This example illustrates a low-noise situation
where differences are less obvious. The fine detail throughout the
bone is better captured by offset-1 sampling, compared to the highlighted
region of the offset-0 version. There are clear differences in detail
shown in the central high-lighted region.}
\end{figure}

\begin{figure}
\subfloat[Fully sampled ground truth]{\includegraphics[width=0.49\columnwidth]{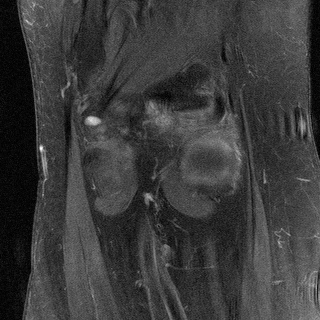}

}\subfloat[Offset 0 sampling ]{'\includegraphics[width=0.49\columnwidth]{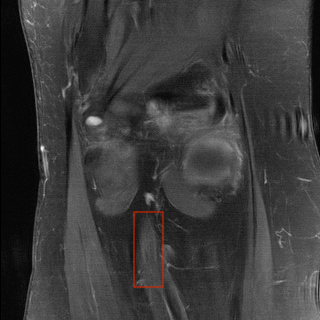}

}

\subfloat[Offset 1 sampling]{\includegraphics[width=0.49\columnwidth]{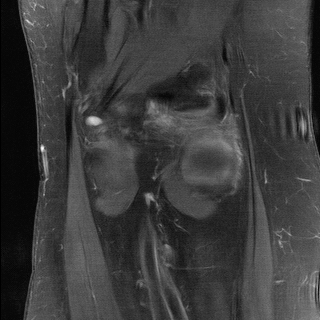}

}\subfloat[Random sampling ]{\includegraphics[width=0.49\columnwidth]{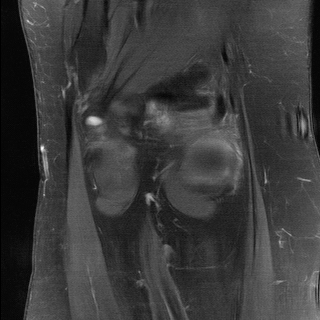}

}\caption{\label{fig:examples-1-1-1}This example illustrates the larger differences
seen in high-noise situations. The offset-1 sampling mask retains
large-scale structure compared to the offset-0 sampling mask in the
highlighted region.}
\end{figure}

\end{document}